\documentstyle[floats,preprint,aps,epsf]{revtex}
\tightenlines
\begin{document}
	
\title{A new time-machine model with compact vacuum core}

\author{Amos Ori}

\address{Department of Physics,\\
Technion---Israel Institute of Technology, Haifa, 32000, Israel}
\date{\today}

\maketitle

\begin{abstract}
We present a class of curved-spacetime vacuum solutions which develope
closed timelike curves at some particular moment. We then use these vacuum
solutions to construct a time-machine model. The causality violation occurs
inside an empty torus, which constitutes the time-machine core. The matter
field surrounding this empty torus satisfies the weak, dominant, and strong
energy conditions. The model is regular, asymptotically-flat, and
topologically-trivial. Stability remains the main open question.
\end{abstract}
\vspace{2ex}
\vspace{2ex}

The problem of time-machine formation is one of the outstanding open
questions in spacetime physics. Time machines are spacetime configurations
including closed timelike curves (CTCs), allowing physical observes to
return to their own past. In the presence of a time machine, our usual
notion of causality does not hold. The main question is: Do the laws of
nature allow, in principle, the creation of a time machine from ''normal''
initial conditions? (By ''normal'' I mean, in particular, initial state with
no CTCs.)

Several types of time-machine models were explored so far. The early
proposals include Godel's rotating-dust cosmological model \cite{1} and
Tipler's rotating-string solution \cite{2}. More modern proposals include
the wormhole model by Moris, Thorne and Yurtsever \cite{3}, and Gott's
solution \cite{4} of two infinitely-long cosmic strings. Ori \cite{5} later
presented a time-machine model which is asymptotically-flat and
topologically-trivial. Later Alcubierre introduced the warp-drive concept 
\cite{6}, which may also lead to CTCs.

The above models, however, all suffer from one or more severe problems. In
some of them \cite{3}\cite{6} the weak energy condition (WEC) \cite{7} is
violated, indicating unrealistic matter-energy content. The WEC states that
for any physical (timelike) observer the energy density is non-negative,
which is the case for all known types of (classical) matter fields. In other
models the CTCs are either pre-existing \cite{1}\cite{2}\cite{4} and/or
''come from infinity'' \cite{4} (see \cite{14}\cite{13}), and/or there is a
curvature singularity \cite{2}.

The only of the above models which does not violate the WEC, and in which
CTCs evolve from normal initial data (and within a compact region of space)
is that of Ref. \cite{5}. But this model, too, is not satisfactory, for the
following reason: The energy-momentum source occupying the time-machine
core, though consistent with the WEC (and also with the dominant energy
condition \cite{5a}), does not fit any known type of matter field. Therefore
this configuration (spacetime plus matter) is not a solution of any
prescribed set of field equations. Even if we assumed there exist physical
matter fields that yield the desired initial configuration, we cannot tell
how these fields (and the geometry) will evolve in time, because we do not
have at hand the set of evolution equations. It therefore leaves open the
question of whether the system will or will not form CTCs. All we can say is
that the WEC does not preclude the formation of CTCs.

We would therefor like a good time-machine model to be made of a well known
matter field, preferably an elementary one. Obviously the most elementary
field (in the present context) is the pure gravitational field, i.e. a
vacuum spacetime. It is primarily this issue which we address in the present
paper.

There are several other desired features which we would like our model to
satisfy: We want the spacetime (and particularly the initial hypersurface)
to be asymptotically-flat (or, alternatively, asymptotically-Friedmann). In
addition, we would like the onset of causality violation to take place
within a {\it finite} region of space. That is, we would like the initial
hypersurface to include a compact piece $S_{0}$, such that the onset of
causality violation---the appearance of a closed causal loop---will be fully
dictated by the initial conditions on $S_{0}$. This requirement makes sense,
because presumably even an advanced civilization will not be able to control
the initial data in the entire space, but only in a finite region (in the
best case). Note that this criterion for compactness differs from the notion
of ''compact generation'' introduced by Hawking \cite{9}, as we further
discuss below. We shall refer to the compact region of spacetime including 
$S_{0}$ and also its future domain of dependence and its neighborhood, as the 
{\it core} of the time-machine spacetime (in our model this will include the
first causal loop and the CTCs in its immediate neighborhood).

In this paper we present a class of vacuum solutions in which CTCs form at
some particular moment. We then use such a vacuum solution as the core of a
new type of time-machine model which satisfies all the above requirements.
In this model the evolution starts from a regular initial spacelike
hypersurface (partial Cauchy surface) which (like spacetime itself) is
asymptotically flat, topologically-trivial, and satisfies the weak,
dominant, and strong energy conditions (the {\it energy conditions}). Then
CTCs form in the central region at some particular moment.

Our model consists of an empty (i.e. vacuum) 
torus which constitutes the time-machine
core. This toroidal vacuum region is immersed in a larger, sphere-like,
region of matter satisfying the energy conditions. The matter region is
finite, and is surrounded by an external asymptotically-flat vacuum region
(specifically in our construction it is the Schwarzschild geometry). The
matter in the intermediate range, though satisfying the energy conditions,
is not associated as yet with any specific known matter field (though it
appears likely that it will be possible to compose it from e.g. some
combination of neutral and charged dusts, and electromagnetic fields); Hence
we cannot tell yet what is the matter's evolution equation. Nevertheless,
the causality violation occurs in the internal vacuum core, and it is
completely dictated by the initial conditions at the compact set $S_{0}$
(located in the vacuum core too). Here we shall primarily discuss the vacuum
solution at the time-machine core (which is the main new feature in this
paper); the structure of the surrounding matter and its matching to the
external asymptotically-flat universe will be presented elsewhere \cite{full}.

There are two rather general analyses, by Tipler \cite{8} and by Hawking 
\cite{9}, which put constraints on the creation of a time machine in a
compact region of space without violating the WEC. At first sight each of
these analyses might appear to preclude a model like the one presented here.
A closer look, however, reveals that there is no inconsistency. This was
already demonstrated and explained in Ref. \cite{5} (which, too, presents a
time-machine model satisfying the WEC). In short, Hawking's analysis refers
to {\it compactly-generated} time machines, which probably is not the case
here (see below). Our model is consistent with Tipler's analysis because it
includes a closed null geodesic (denoted N below) which is future-incomplete
(though no local irregularity occurs there; See also the discussion in Ref. 
\cite{5}).

We turn now to describe the geometry of the time-machine core. Consider the
vacuum solution 
\begin{equation}
ds^{2}=dx^{2}+dy^{2}-2dzdT+[f(x,y,z)-T]dz^{2}.  \label{general}
\end{equation}
The coordinates $(x,y,T)$ get all real values (though we later truncate the
solution in $x$ and $y$ at the internal boundary of the matter region), but 
$z$ is a cyclic coordinate, $0\leq z\leq L$ for some $L>0$, with $z=L$ and 
$z=0$ identified. $f$ is any function (properly periodic in $z$) satisfying 
\footnote{%
For $f$ not satisfying Eq. (\ref{vacuum}) the metric (\ref{general})
represents null dust with $G_{zz}$ given by $-(1/2)(f_{,xx}+f_{,yy})$ and
all other components vanishing. We shall not discuss this case here.} 
\begin{equation}
f_{,xx}+f_{,yy}=0.  \label{vacuum}
\end{equation}
It can be shown that this class is locally isometric to vacuum plane-fronted
waves \cite{exact}. \footnote{%
We should mention here two simple extensions of the vacuum solution (\ref
{general}). First, replace $g_{zz}$ therein by $f(x,y,z)-h(z)T$, for any
periodic function $h(z)$ [this extension is globally-nontrivial for any
sign-changing function $h(z)$]. Second, to the metric (\ref{general}) add 
$g_{xz}=wy$, $g_{yz}=-wx$, and also add $w^{2}(x^{2}+y^{2})$ to $g_{zz}$, for
any $w\neq 0$.} However, its global properties are completely different from
those of plane-fronted waves, as we now discuss.

One immediately observes that the metric (\ref{general}) develops CTCs at
sufficiently large $T$. For each $x,y$, the metric function 
$g_{zz}=f(x,y,z)-T$ is positive (for all $z$) at small $T$, but becomes
negative (for all $z$) at sufficiently large $T$. Consequently the closed
curves of constant $x,y,T$ are spacelike at small $T$ but become
timelike---namely CTCs---at large $T$. (Below we consider a specific example
in more detail.) Note that the metric is everywhere regular with 
$\det(g)=-1$, so no local pathology is involved in the formation of CTCs.

For generic $f$ the spacetime is curved, with 
\begin{equation}
R_{izjz}=-(1/2)f_{,ij}
\end{equation}
(and its obvious permutations), and all other components vanishing, where 
$i,j$ stand for $x$ and $y$. The metric becomes locally-flat in the
degenerate case $f=0$ [the same holds for any $f=f(z)$]. This is just the
Misner space \cite{12}, generalized to four dimensions in a straightforward
manner. The degenerate $f=0$ spacetime shares some of the features of the
generic-$f$ non-degenerate case, though its global structure is different
and more pathological (see below).

For concreteness we now specialize to a simple example. We take 
\begin{equation}
f=a(x^{2}-y^{2})/2  \label{specific}
\end{equation}
for some constant $a>0$. This yields an empty curved spacetime, locally
isometric to a linearly-polarized plane wave. We now transform from $T$ to a
new time coordinate 
\[
t=T-a(x^{2}-y^{2})/2+e\rho ^{2} 
\]
for some $e>0$, where $\rho ^{2}\equiv x^{2}+y^{2}$. This turns out to be
convenient, because the hypersurfaces $t=const$ provide a useful foliation
for the analysis below. The line element becomes 
\begin{equation}
ds^{2}=dx^{2}+dy^{2}-2dzdt+(e\rho ^{2}-t)dz^{2}+2[(2e-a)xdx+(2e+a)ydy]dz.
\label{simp}
\end{equation}
This metric is empty and curved (as before), and again one finds 
$\det(g)=-1 $. 
Since $g_{zz}=e\rho ^{2}-t$, the closed curves of constant $x,y,t$
are spacelike at $t<e\rho ^{2}$ and timelike at $t>e\rho ^{2}$ (and null at 
$t=e\rho ^{2}$). These curves are non-geodesic, except for the single closed
null geodesic $x=y=t=0$, which we denote N.

A hypersurface $t=const$ is spacelike when $g^{tt}<0$ and timelike whenever 
$g^{tt}>0$. One finds 
\[
g^{tt}=t+(2e-a)^{2}x^{2}+(2e+a)^{2}y^{2}-e\rho ^{2}. 
\]
Choose now sufficiently small $a>0$ and $e>a$ such that $e>(2e+a)^{2}$. Then
the hypersurfaces $t=const$ may be characterized as follows:

(I) For $t<0$ they are spacelike throughout;

(II) The hypersurface $t=0$ is spacelike everywhere, except at the central
curve, $x=y=0$ (the geodesic N), where it is null ($g^{tt}=0$).

(For $t>0$ the hypersurfaces $t=const$ are mixed: timelike for sufficiently
small $\rho $ and spacelike at large $\rho $.)

In the degenerate case $f=0$ (the 4d Misner space), the hypersurface $T=0$
is null, and all its generators are closed null geodesics (the curves of
constant $x,y$). This case is pathological, because the analytic extension
beyond $T=0$ is non-unique \cite{12}. No such pathology occurs in the
non-degenerate case. For a generic $f$, the closed null geodesics are
isolated [like N in the specific example (\ref{specific})] and do not 
form a
hypersurface. Note also that in the degenerate case $f=0$ the causal
structure is inherently unstable in the following sense: Adding an
arbitrarily-small fixed number to $g_{xz}$ and/or $g_{yz}$ (which vanish
otherwise) leaves the geometry vacuum and locally-flat, but the hypersurface
of closed null geodesics entirely disappears. (This metric is closely
related to the Grant-Gott spacetime \cite{13}, which includes no closed
causal geodesics.) Such a global instability does not occur in the
non-degenerate case, which we consider throughout this paper.

The above constructed vacuum solution is now used as the core of our
time-machine spacetime, with $x=y=0$ located at the central circle of the
torus. The solution (\ref{simp}) is truncated at $\rho =\rho _{0}$ for some 
$\rho _{0}>0$, where the matter region starts.

As was discussed above, we would like the region of CTCs to evolve, in a
deterministic manner, from a compact spacelike hypersurface $S_{0}$ in the
vacuum core. This would be fully accomplished if a region of CTCs were
included in $D^{+}(S_{0})$ ($D^{+}$ denotes the future domain of dependence 
\cite{7}, also known as the future Cauchy development). But obviously this
can never be the case because by definition $D^{+}$ of any spacelike
hypersurface cannot include any closed causal curve. It may be possible,
however, that the {\it boundary} of $D^{+}(S_{0})$ will include a closed
causal curve (in fact a closed null geodesic); And this appears to be the
maximum one can hope for, in terms of the causal relation between $S_{0}$
and the region of causality violation. We shall now show that our model
indeed has this desired feature, namely, one can choose a compact spacelike
hypersurface $S_{0}$ in the vacuum core, such that a closed causal curve
(the null geodesic N) appears at the boundary of $D^{+}(S_{0})$.
Furthermore, any regular extension of the geometry beyond N will include a
region of CTCs.

For any $\rho _{1}>0$ the hypersurfaces $t=const\leq 0$ are all spacelike at 
$\rho \geq \rho _{1}$. Also, the hypersurfaces $t=const<0$ are spacelike
even at $\rho <\rho _{1}$. Any such spacelike hypersurface may be slightly
deformed and still remain spacelike. Consider, in particular, the composed
hypersurface given by $t=const=t_{0}<0$ at $\rho \leq \rho _{1}$, by $t=0$
at $\rho \geq \rho _{2}$, and by some interpolating function $t=\tilde{t}
(\rho )$ in the range $\rho _{1}\leq \rho \leq \rho _{2}$, for some
parameters $0<\rho _{1}<\rho _{2}<\rho _{0}$. The function $\tilde{t}(\rho )$
is chosen to be a monotonic one, which smoothly bridges between $t=t_{0}$ at 
$\rho =\rho _{1}$ and $t=0$ at $\rho =\rho _{2}$. For sufficiently small 
$|t_{0}|$, the function $\tilde{t}(\rho )$ 
may be taken to be of sufficiently
small slope, such that the hypersurface described by $t=\tilde{t}(\rho )$ is
spacelike in $\rho _{1}\leq \rho \leq \rho _{2}$. By this we have
constructed a composed spacelike hypersurface in the entire range $\rho \leq
\rho _{0}$. We then restrict this hypersurface to the range $\rho \leq \rho
_{3}$ for some $\rho _{2}<\rho _{3}<\rho _{0}$, and denote this compact
spacelike hypersurface by $S_{0}$.

Take now any point P at $t<0$ in a sufficiently small neighborhood of N
(located at $\rho =t=0$), and consider any inextendible past-directed causal
curve $\gamma $ emanating from P. (In particular P is located at $\rho <\rho
_{1}$ and $t>t_{0}$, i.e. at the future of $S_{0}$.) Since in the range 
$t_{0}\leq t<0$ all hypersurfaces $t=const$ are entirely spacelike, one can
easily show that $\gamma $ has no other choice but to move towards smaller $t
$ values, until it intersects $S_{0}$. Therefore the point P belongs to 
$D^{+}(S_{0})$. This means that N resides at the boundary of $D^{+}(S_{0})$.

The compact initial hypersurface $S_{0}$ can be extended through the matter
region and the external vacuum region to spacelike infinity, to form a
regular, asymptotically-flat, partial Cauchy surface, everywhere satisfying
the energy conditions, which we denote $\Sigma $. This will be demonstrated
elsewhere. \cite{full}

We conclude that in our model the onset of causality violation---the
appearance of the closed null geodesic N---is fully dictated by the initial
conditions in the compact vacuum piece $S_{0}$ of the initial hypersurface 
$\Sigma $. In particular, the details of the initial data at the external
vacuum region, or at the intermediate matter region, cannot interfere with
this onset of causality violation. Although the above construction only
demonstrates the inevitable formation of a single closed causal orbit, 
{\it any} smooth ($C^{1}$ metric) extension of the 
geometry at N to $t>0$ will
include a region of CTCs, because $g_{zz,t}<0$ at N.

Hawking \cite{9} previously introduced another notion of compactness: A 
{\it compactly generated} time machine is 
one in which all null generators of the
chronology horizon, when traced back to the past, enter a compact region of
spacetime and remain there. He then showed that a compactly-generated time
machine (with non-compact initial hypersurface) must include a region where
the WEC is violated. Our model evades this WEC violation, probably because
it is not compactly generated. In the intermediate matter region, the
initial data for the metric are constructed to be regular and smooth (and to
satisfy the WEC) \cite{full}. This holds on $\Sigma $ and also for some time
interval to its future. But our construction by no means guarantees that the
regular WEC-satisfying metric at the matter region can extend to all future
times. From Hawking's arguments \cite{9} it appears to follow that the
matter geometry, when forced to satisfy the WEC, must develope some
non-compactness (e.g. a singularity or an ''internal infinity'') at later
times. Then these arguments, applied to our construction, may further
suggest that generators of the chronology horizon will emanate from this
non-compact region. But this proposed scenario must be verified by an
explicit study of the structure of the Cauchy horizon. We emphasize again
that although probably not ''compactly generated'', our model does
demonstrate the formation of closed causal loops from the initial data on a
compact vacuum region $S_{0}$. Although it is only a single closed null
geodesic which resides in the closure of $D^{+}(S_{0})$, any smooth ($C^{1}$
) extension of the metric to $t>0$ will include a region of CTCs.

The fact that the weak, dominant, and strong energy conditions are satisfied
suggests that the initial conditions required for our model are physically
acceptable. This does not mean that we shall be able to practically initiate
such initial conditions in the foreseeable future. However, perhaps an
advanced civilization will be able to do this (and perhaps even natural
processes, involving large gravitating masses in rapid motion, may lead to
such conditions). There still remains, however, the issue of {\it stability}
. Several analyses \cite{9}\cite{9a}\cite{9b} indicated possible
instabilities of various time-machine solutions to classical perturbations
and/or quantum-mechanical fluctuations (see however \cite{10}). Whereas
these analyses mostly referred to compactly-generated models, some of the
arguments for quantum instabilities apply to noncompactly-generated models
as well (see in particular the discussion in \cite{9b}). These instabilities
may raise doubts on whether a model like the one presented here can be
implemented in reality. Yet, it appears that so far these indications for
instability do not rule out the possibility of actual time-machine
construction. The strength of the quantum instability is not clear yet, and,
more importantly, it is not known yet what will be the {\it outcome} of this
instability (namely, what will be the spacetime configuration that
eventually forms). Perhaps we shall have to await the formulation of the
full theory of quantum gravity before we know whether quantum instabilities
provide chronology protection or not (see discussion in \cite{15}). The
situation with regards of classical instabilities is somewhat different:
Their occurrence still needs be established, especially in the
noncompactly-generated case and beyond the context of geometrical optics. If
classical instabilities are found to be inevitable, their strength and their
outcome can be explored by evolving the classical Einstein equations for
slightly-perturbed initial conditions.

\end{document}